\documentclass[aps,prl,preprint]{revtex4-1}

\usepackage{amsmath}
\usepackage{slashed}

\newcommand{\rf}[1]{(\ref{#1})}
\def\etal{{\it et al.}}

\usepackage{xspace}
\usepackage{relsize}
\def\babar{\mbox{\slshape B\kern-0.1em{\smaller A}\kern-0.1em
    B\kern-0.1em{\smaller A\kern-0.2em R}}\xspace}

\begin{document}

\title{Electromagnetic flavor-changing lepton decays 
\\
from Lorentz and CPT violation}

\author{V.\ Alan Kosteleck\'y,$^1$ E.\ Passemar,$^1$ and N.\ Sherrill$^2$}

\affiliation{$^1$Physics Department, Indiana University, 
Bloomington, Indiana 47405, USA\\
$^2$Department of Physics and Astronomy, University of Sussex, 
\\
Brighton BN1 9QH, UK
}

\date{July 2022}

\begin{abstract}
Lorentz- and CPT-violating effects 
initiating two-body electromagnetic flavor-changing decays of charged leptons 
are studied in the framework of Lorentz-violating effective field theory.
An analysis of data from experiments 
at the Paul Scherrer Institute and at the Stanford Linear Accelerator 
measuring the branching ratios of these decays
provides 576 constraints 
on independent flavor-changing effects in the charged-lepton sector,
consistent with no Lorentz and CPT violation at the level
of parts in $10^{-13}$~GeV$^{-1}$ to $10^{-9}$~GeV$^{-1}$.
\end{abstract}

\maketitle

For many decades now,
flavor-changing effects have played a central role
in the discovery of new physics violating fundamental symmetries of nature,
including the discrete symmetries
charge conjugation C, parity inversion P, and time reversal T
and the continuous internal symmetries of the minimal Standard Model (SM). 
In the SM,
for example, 
charged weak interactions change fermion flavor,
converting charged leptons to neutrinos
or mixing quarks of different flavors.
These effects underlie the observation of P violation in weak decays
\cite{wu57}
and the detection of CP violation in kaon oscillations
\cite{jc64}.
Also,
flavor oscillations of neutrinos
have provided evidence of physics beyond the SM
\cite{rd68,sk98},
involving breaking of the accidental SM global 
U$_e$(1)$\times $U$_\mu$(1)$ \times $U$_\tau$(1)
invariance by right-handed neutrino fields.

A key symmetry in nature is Lorentz invariance,
which ensures that physical laws are unchanged under rotations or boosts
and is accompanied by CPT invariance.
While these invariances hold to an excellent approximation,
they could be broken in an underlying theory 
that combines gravity and quantum physics
such as strings
\cite{ksp},
thereby leading to tiny observable effects 
of Lorentz violation (LV) at present energy scales.
Although extensive experimental investigations of this idea 
have been performed 
\cite{tables},
comparatively little is known about flavor-changing LV interactions.
Instead,
most studies of flavor-changing LV effects involve propagation. 
For example,
neutral-meson oscillations are sensitive to Lorentz- and CPT-violating effects
that are otherwise challenging to detect
\cite{ak98},
but in these experiments the flavor changes
are driven by known weak interactions
while the Lorentz and CPT violation is diagonal in quark flavor.

The present work addresses this gap in the literature
by investigating flavor-changing LV interactions 
that induce charged-lepton decays,
in particular electromagnetic decays of the muon and tau.
In conventional Lorentz-invariant models,
flavor-changing decays of charged leptons occur 
only via suppressed one-loop processes
with branching ratios $\lesssim10^{-54}$
\cite{smdecays},
so these processes offer exceptionally clean probes of new physics.
Here,
we perform a model-independent analysis of experimental data 
to search for dominant LV effects in these decays.
Our results are consistent with no effects 
in 576 independent coefficients for LV,
thereby excluding electromagnetic flavor-changing LV interactions 
of leptons at parts in $10^{-13}$~GeV$^{-1}$ to $10^{-9}$~GeV$^{-1}$.

Given the absence of compelling evidence for LV to date,
model-independent techniques are desirable and appropriate
for analyses of prospective low-energy signals.
A model-independent framework based on effective field theory,
known as the Standard-Model Extension (SME) 
\cite{ck97,ak04},
offers a powerful and widely adopted approach
for experimental searches for LV
\cite{tables,review}.
In Minkowski spacetime,
the Lagrange density of the SME contains the SM
extended by adding all observer-invariant terms 
formed by contracting LV operators with controlling coefficients.
In effective field theory,
CPT violation implies LV 
\cite{ck97,owg}, 
so the SME also describes all CPT-violating effects.
Despite the substantial body of existing experimental measurements
\cite{tables},
many coefficients for LV remain unconstrained to date.
Their magnitudes are generically undetermined by theory,
with some ``countershaded'' ones challenging to detect despite being large 
\cite{kt09},
so model-independent experimental searches 
without prior assumptions about coefficient magnitudes
acquire particular importance in this context.

\renewcommand\arraystretch{1.6}
\begin{table*}
\caption{
\label{operators}
Dimension-five terms with $F_{\mu\nu}$ couplings.
Note $(-)^\mu  \equiv +$ for $\mu = 0$
and $(-)^\mu \equiv -$ for $\mu = 1,2,3$.}
\centering 
\setlength{\tabcolsep}{4pt}
\begin{tabular}{c c c c c c c} 
\hline\hline
Term & No. & C & P & T & CP & CPT \\ 
\hline 
$	-\tfrac{1}{2}(m_{F}^{(5)})^{\alpha\beta}_{AB}F_{\alpha\beta}\overline{\psi}_A\psi_B	$&$	54	$&$	-	$&$	(-)^\alpha (-)^\beta	$&$	-(-)^\alpha (-)^\beta	$&$	-(-)^\alpha (-)^\beta	$&$	+	$	\\
$	- \tfrac{1}{2}i(m_{5F}^{(5)})^{\alpha\beta}_{AB}F_{\alpha\beta}\overline{\psi}_A\gamma_5\psi_B	$&$	54	$&$	-	$&$	-(-)^\alpha (-)^\beta	$&$	(-)^\alpha (-)^\beta	$&$	(-)^\alpha (-)^\beta	$&$	+	$	\\
$	- \tfrac{1}{2}(a_{F}^{(5)})^{\mu\alpha\beta}_{AB}F_{\alpha\beta}\overline{\psi}_A\gamma_\mu\psi_B	$&$	216	$&$	+	$&$	(-)^\mu (-)^\alpha (-)^\beta	$&$	-(-)^\mu (-)^\alpha (-)^\beta	$&$	(-)^\mu (-)^\alpha (-)^\beta	$&$	-	$	\\
$	- \tfrac{1}{2}(b_{F}^{(5)})^{\mu\alpha\beta}_{AB}F_{\alpha\beta}\overline{\psi}_A\gamma_\mu\gamma_5\psi_B	$&$	216	$&$	-	$&$	-(-)^\mu (-)^\alpha (-)^\beta	$&$	-(-)^\mu (-)^\alpha (-)^\beta	$&$	(-)^\mu (-)^\alpha (-)^\beta	$&$	-	$	\\
$	- \tfrac{1}{4}(H_{F}^{(5)})^{\mu\nu\alpha\beta}_{AB}F_{\alpha\beta}\overline{\psi}_A\sigma_{\mu\nu}\psi_B	$&$	324	$&$	+	$&$	(-)^\mu(-)^\nu (-)^\alpha (-)^\beta	$&$	(-)^\mu(-)^\nu (-)^\alpha (-)^\beta	$&$	(-)^\mu(-)^\nu (-)^\alpha (-)^\beta	$&$	+	$	\\
\hline\hline 
\end{tabular}
\end{table*}

The presence of LV allows the electromagnetic decays 
$\ell_A \rightarrow \ell_B + \gamma$
of a charged lepton $\ell_A$ 
into a charged lepton $\ell_B$ and a photon $\gamma$
to proceed directly at tree level,
in contrast to the suppressed loop-level decays
in conventional Lorentz-invariant models.
Since the decays are governed by energy scales on the order of $m_A$,
SME operators of low mass dimension are expected to provide 
the dominant experimental signals in these and related processes
\cite{ck97,cg99,yi03,gh20,cks20,hp11,nowt16,lmt16,msf17,gmn20}.
All terms in the SME Lagrange density 
with operators of mass dimension up to six are explicitly known 
\cite{kl19}.
In the lepton-photon sector,
some of these operators affect propagation
while others represent pure interactions.
The former involve bilinears in the lepton fields 
and their spacetime derivatives.
Setting the photon field to zero
establishes the free-fermion Lagrange density
and determines the propagating flavor eigenstates of the hamiltonian.
By construction,
these eigenstates represent the physical electron, muon, and tau fields
relevant for laboratory experiments,
and so during free propagation
they preserve the corresponding lepton numbers without flavor changes
despite the presence of LV.
With the photon reintroduced, 
the spacetime derivatives in all operators affecting propagation 
are covariant and symmetrized
\cite{kl19},
so the photon fields appearing in these bilinears
also preserve the eigenstates during propagation.
It follows that no tree-level electromagnetic flavor-changing decays 
of the physical charged leptons can occur
from these covariant-derivative couplings,
contrary to the assumptions of earlier works on these decays 
that adopted experimentally unphysical eigenstates for calculations.
This flavor-conserving feature has the same origin as its analogue in the SM,
and it can be understood as a consequence of the global symmetry 
U$_e$(1)$\times $U$_\mu$(1)$ \times $U$_\tau$(1)
of the free-fermion LV theory that is transmitted 
to any photon couplings associated with covariant derivatives.

Since we are interested in operators
that change the flavor of a physical eigenstate while emitting a photon,
the effects relevant here must instead involve 
direct couplings of the electromagnetic field strength to lepton bilinears.
All such operators are independent of the propagation terms 
in the Lagrange density,
so they can be off-diagonal in flavor space 
even in the basis of physical eigenstates relevant for experiments.
They therefore can violate the 
U$_e$(1)$\times $U$_\mu$(1)$ \times $U$_\tau$(1) symmetry,
inducing observable charged-lepton decays in detectors. 
The dominant operators of this form have mass dimension five
and are the focus of this work. 
Since the lepton decay rates contain the square of the decay amplitude,
which itself is already at leading order in LV,
any LV effects in propagation
and the associated modifications of phase-space factors 
can be disregarded in what follows. 

The dimension-five terms of interest in the Lagrange density 
\cite{kl19}
are listed in the first column of 
Table \ref{operators}. 
The lepton fields are denoted $\psi_A$, $A = e,\mu,\tau$,
and the electromagnetic field strength is $F_{\mu\nu}$.
The behaviors of the terms
under C, P, T, CP, and CPT
are also displayed in the table. 
The results reveal the prospect of novel sources
of discrete-symmetry breakdown in the presence of LV, 
including possible violations of the CPT theorem
\cite{cpt}.
The coefficients for LV 
$(m_{F}^{(5)})^{\alpha\beta}_{AB}$, 
$(m_{5F}^{(5)})^{\alpha\beta}_{AB}$, 
$(a_{F}^{(5)})^{\mu\alpha\beta}_{AB}$, 
$(b_{F}^{(5)})^{\mu\alpha\beta}_{AB}$, 
$(H_{F}^{(5)})^{\mu\nu\alpha\beta}_{AB}$
have units of GeV$^{-1}$ 
and by construction are antisymmetric 
on the index pairs $(\alpha,\beta)$ and $(\mu,\nu)$. 
They can be viewed as complex matrices in flavor space, 
constrained by hermiticity of the Lagrange density.
The number of independent real components of each coefficient
is given in the second column of the table.

The SME coefficients transform as covariant tensors 
under observer Lorentz transformations
and as scalars under particle transformations,
so the terms in the first column generically violate Lorentz invariance
\cite{ck97}.
However,
some of the coefficients can contain components proportional 
to products of the Minkowski metric $\eta_{\mu\nu}$
and the Levi-Civita tensor $\epsilon_{\mu\nu\rho\sigma}$,
which are Lorentz-group invariants,
and hence the corresponding terms are Lorentz invariant.
For example,
the coefficients
$(H_{F}^{(5)})^{\mu\nu\alpha\beta}_{AB}$
contain two Lorentz-invariant pieces yielding the Lorentz-invariant terms 
$-(H_{F,1}^{(5)})_{AB}F^{\mu\nu}
\overline\psi_A\sigma_{\mu\nu}\psi_B$
and  
$- (H_{F,2}^{(5)})_{AB}\widetilde{F}^{\mu\nu}
\overline\psi_A\sigma_{\mu\nu}\psi_B$,
where 
$\widetilde{F}^{\mu\nu} \equiv \epsilon^{\mu\nu\alpha\beta}F_{\alpha\beta}/2$
is the dual field strength.
These terms describe anomalous magnetic and electric dipole moments.
They correspond to the leading operators 
in the Low-energy Effective Field Theory (LEFT)
\cite{nonpertmue,taudecays}
for the decay $\ell_A \rightarrow \ell_B + \gamma$,
which derive from dimension-six effects in the 
Standard-Model Effective Field Theory (SMEFT) 
upon matching operators at the electroweak scale
\cite{EFTmatching}.
For the magnetic dipole term in the two-flavor electron-muon limit,
for instance,
the connection appears explicitly by expanding the SME results
into chiral field components
and matching to the LEFT and SMEFT operators. 
More generally,
the SMEFT can be viewed as a restriction of the SME
to the Lorentz-invariant sector in Minkowski spacetime,
SME~$\supset$~SMEFT~$\supset$~LEFT,
with every SMEFT Wilson coefficient
being a Lorentz-invariant combination
of nonminimal SME coefficients in Minkowski spacetime
and an appropriate power of the high-energy scale
representing the onset of new physics.

The terms listed in Table \ref{operators}
generate novel three-point vertex functions allowing the decays
$\ell_A \rightarrow \ell_B + \gamma$.
These terms leave unaffected the free propagation 
of the fermion and photon fields,
so standard quantization techniques apply
\cite{kl01}.
If $\ell_A$ has momentum $p_\mu$ and spin projection $s$,
$\ell_B$ has momentum $p'_\mu$ and spin projection $s'$,
and the photon has momentum $k_\mu$ and helicity $\lambda$,
then the contribution to the decay amplitude 
$\mathcal{M}_{AB}^{(s,s',\lambda)} (p,p',k)$
from a given term in the table takes the form 
\begin{align}
\mathcal{M}_{AB}^{(s,s',\lambda)} = 
\begin{cases}  
\overline{u}_B^{(s)}(p')V^{\beta}_{BA}(k)u_A^{(s')}(p)
\epsilon^{(\lambda)*}_\beta(k),
\\
\overline{v}_A^{(s)}(p)V^{\beta}_{AB}(k)v_B^{(s')}(p')
\epsilon^{(\lambda)*}_\beta(k),
\end{cases}
\label{gendecayamp}
\end{align}
where the first line holds for particle decay 
and the second for antiparticle decay.
The quantity $V^{\beta}_{AB}(k)=(V^{\beta}_{BA}(k))^*$
is the member of the set 
$\{(m_{F}^{(5)})^{\alpha\beta}_{AB} k_\alpha$,
$i(m_{5F}^{(5)})^{\alpha\beta}_{AB}\gamma_5 k_\alpha$,
$(a_{F}^{(5)})^{\mu\alpha\beta}_{AB}\gamma_\mu k_\alpha$,
$(b_{F}^{(5)})^{\mu\alpha\beta}_{AB}\gamma_\mu\gamma_5 k_\alpha$,
$\tfrac{1}{2}(H_{F}^{(5)})^{\mu\nu\alpha\beta}_{AB}\sigma_{\mu\nu} k_\alpha\}$
corresponding to the chosen term in Table \ref{operators}.
Note that the Ward identity ensuring gauge invariance 
of the amplitude \eqref{gendecayamp} 
is enforced through the vanishing of
$\mathcal{M}_{AB}^{(s,s',\lambda)} (p,p',k)$
under the replacement $\epsilon^{(\lambda)*}_\beta(k) \rightarrow k_\beta$.

Existing experimental limits on charged-lepton transitions 
are stringent and hence well suited for constraining
small deviations from known physics. 
Tight bounds on flavor violations involving muons 
come from studies of two-body decays 
by the Mu to Electron Gamma (MEG) collaboration
at the Paul Scherrer Institute
\cite{MEGlimit},
BR$(\mu^+ \rightarrow e^+ +\gamma) 
\leq 4.2\times 10^{-13}$.
The \babar collaboration 
at the Stanford Linear Accelerator 
obtained constraints 
\cite{BABARlimit}
both on decays of taus into muons,
BR$(\tau^\pm \rightarrow \mu^\pm +\gamma) 
\leq 4.4\times 10^{-8}$,
and on decays into electrons,
BR$(\tau^\pm \rightarrow e^\pm +\gamma) 
\leq 3.3\times 10^{-8}$. 

In the MEG experiment
\cite{MEGdetector}, 
polarized antimuons in a beam are stopped in a plastic target
and subsequently decay at rest,
producing back-to-back positrons and photons 
each carrying energy $m_\mu/2 \simeq 52.83$ MeV.
The signal process therefore involves the calculation
of the integrated decay rate of a polarized antimuon at rest 
to a positron and photon.
The direction dependence arising from LV 
means that the calculation must take into account 
the restricted solid-angle acceptance window for the signal photon
in the MEG detector
and allow for all possible spin configurations in the final state.
It is convenient to identify the beam direction as the detector $z$ axis.
Approximately $\simeq 11\%$ of the full phase space is accessible, 
and the limits on detector polar and azimuthal angles are 
$\theta \in (1.21, 1.93)$, 
$\phi \in (\tfrac{2\pi}{3},\tfrac{4\pi}{3})$.
For our purposes it suffices to approximate the antimuons
as having initial polarization $P_\mu = -1$.
In practice a small depolarization of the beam occurs during propagation,
which could be taken into account in a future detailed data reconstruction.
The polarized decay rate is therefore given by 
\begin{align}
\Gamma \approx 
\frac{1}{64\pi^2m_\mu}
\int^{\theta_{\text{max}}}_{\theta_{\text{min}}}
\int^{\phi_{\text{max}}}_{\phi_{\text{min}}} 
\sin\theta d\theta d\phi 
~|\mathcal{M}_{\mu e}(\theta,\phi)|^2,
\label{decayrate}
\end{align}
where the antiparticle decay amplitude \eqref{gendecayamp} is chosen.
The explicit expression for $|\mathcal{M}_{\mu e}(\theta,\phi)|^2$
is obtained directly from Eq.\ \rf{gendecayamp}
but is lengthy and omitted here.
Multiplying the result by the muon lifetime 
$\tau_\mu \simeq
2.2\times 10^{-6}$ s 
gives the theoretical branching ratio
in terms of the coefficients for LV 
appearing in Table \ref{operators},
expressed in the frame of the MEG detector.

In the \babar experiment
\cite{BABARreport,babar17},
the $\tau^{\pm}$ are produced via unpolarized and asymmetric 
$e^+ e^- \rightarrow \tau^+ \tau^-$ collisions 
near the $\Upsilon(4S)$ resonance.
The emerging tau pairs retain nonzero longitudinal momentum, 
so their rest frame differs from the detector frame.
However, 
the boost factor 
$\gamma \approx \sqrt{s}/(2m_\tau) \simeq \mathcal{O}(1)$
between the two frames is comparatively small and can be disregarded 
in studying LV effects,
so the rest frame of the tau pairs 
can reasonably be taken as the detector frame.
For present purposes it suffices to approximate
the fiducial volume of the \babar detector
as spanning the full 4$\pi$ steradians.
In practice small cones involving $\simeq 10$\% of the volume
along the collider beamline directions are unavailable,
and this could be incorporated into a future data analysis. 
The decay rates for the processes
$\tau^\pm \rightarrow (\mu^\pm, e^\pm) + \gamma$ 
are therefore given by
\begin{align}
\Gamma_{AB}^\pm \approx \frac{1}{64\pi^2m_\tau}
\int_{4\pi}d\Omega 
\sum_{s, s',\lambda}
\tfrac{1}{2}
|\mathcal{M}^{\pm(s,s',\lambda)}_{AB}(\theta,\phi)|^2 ,
\label{decayratetau}
\end{align}
where $\theta,\phi$ are the polar and azimuthal angles of the photon 
with respect to the $+z$ axis of the detector frame,
$A = \tau$ and $B = \mu$ or $e$,
and the signs $\pm$ correspond to the lepton charge in the process.
The full integrand $\overline{|\mathcal{M}(\theta,\phi)|^2}$ is lengthy, 
but if attention is restricted to any given operator
in Table \ref{operators} then it takes the form 
\begin{equation}
\overline{|\mathcal{M}(\theta,\phi)|^2} = 
-\tfrac{1}{2}\eta_{\mu\nu}
\text{Tr} \big[(\slashed{p}\mp m_A)V_{AB}^\mu(\slashed{p'} 
\mp m_B)V_{AB}^{\dagger\nu}\big].
\end{equation}
Explicit evaluation gives
\begin{eqnarray}
\sum\overline{|\mathcal{M}_{m_F^{(5)}}|^2}
&=& 
2(m_A m_B + p\cdot p')
(m_{F}^{(5)})^{k\mu}_{AB}
(m_{F}^{(5)})^{*}_{AB\mu}{}^k, 
\nonumber\\
\sum\overline{|\mathcal{M}_{m_{5F}^{(5)}}|^2} 
&=& 
2(m_Am_B - p\cdot p')
(m_{5F}^{(5)})^{k\mu}_{AB}
(m_{5F}^{(5)})^{*}_{AB\mu}{}^k, 
\nonumber\\
\sum\overline{|\mathcal{M}_{a_F^{(5)}}|^2}
&=& 
2(m_Am_B - p\cdot p')
(a_{F}^{(5)})^{\mu\nu k}_{AB}
(a_{F}^{(5)})^{*}_{AB\mu\nu}{}^k
\nonumber\\
&& - 2\big((a_{F}^{(5)})^{p k \nu}_{AB}
(a_{F}^{(5)})^{*p' k }_{AB}{}_{\nu} 
+ \text{h.c.}\big),
\nonumber\\
\sum\overline{|\mathcal{M}_{b_F^{(5)}}|^2}
&=& 
2(m_Am_B + p\cdot p')
(b_{F}^{(5)})^{\mu\nu k}_{AB}
(b_{F}^{(5)})^{*}_{AB\mu\nu}{}^k
\nonumber\\
&& - 2\big((b_{F}^{(5)})^{p k \nu}_{AB}
(b_{F}^{(5)})^{*p' k}_{AB}{}_\nu
+ \text{h.c.}\big),
\nonumber\\
\sum\overline{|\mathcal{M}_{H_F^{(5)}}|^2} 
&=& 
\tfrac{1}{2}(m_Am_B + p\cdot p')
\nonumber\\
&&
\hskip10pt
\times
(H_{F}^{(5)})^{\mu\nu\alpha k}_{AB}
(H_{F}^{(5)})^{*}_{AB\nu\mu\alpha}{}^k 
\nonumber\\
&& + 2\big((H_{F}^{(5)})^{\mu p k \nu}_{AB}
(H_{F}^{(5)})^{*}_{AB\mu}{}^{p'k}{}_\nu 
+ \text{h.c.}\big),
\qquad
\end{eqnarray}
where the sums are over spins
and indices $p$, $p'$, $k$ represent contraction 
with the corresponding momenta.
Substituting these expressions into the decay rate \eqref{decayratetau} 
and multiplying by the tau lifetime
$\tau_\tau \simeq 2.9\times 10^{-13}$ s
yields the theoretical branching ratio 
in terms of coefficients for LV,
expressed in the detector frame.

The presence of LV means that the explicit values 
of the coefficients listed in Table \ref{operators} are frame dependent,
so experimental results must be reported in a specified frame.
The coefficients can be taken as spacetime constants
in cartesian inertial frames near the Earth
\cite{ak04}.
The canonical frame adopted in the literature 
is the Sun-centered frame (SCF)
with right-handed cartesian coordinates $(T,X,Y,Z)$,
where $T$ is zero at the 2000 vernal equinox, 
the $Z$ axis is parallel to the Earth's rotation axis,
and the $X$ axis points from the Earth to the Sun at $T=0$
\cite{sunframe}.  
The Earth's rotation makes all laboratory frames noninertial
and so the coefficients expressed in the laboratory frame are time dependent, 
oscillating at harmonics of the Earth's sidereal frequency 
${\omega_\oplus}\simeq 2\pi /(23~{\rm h}~56~\min)$ 
\cite{ak98}.
Neglecting the Earth's boost,
the transformation from the SCF to a standard laboratory frame
with $x$ axis pointing to the south, 
$y$ axis to the east, 
and $z$ axis to the local zenith is
\begin{align}
\label{standardrot}
\mathcal{R} =\begin{pmatrix}\cos\chi\cos\omega_{\oplus} T_{\oplus} 
& \cos\chi\sin\omega_{\oplus} T_{\oplus} 
& -\sin\chi \\ -\sin\omega_{\oplus} T_{\oplus} 
& \cos\omega_{\oplus} T_{\oplus} & 0 \\ 
\sin\chi\cos\omega_{\oplus} T_{\oplus} & 
\sin\chi\sin\omega_{\oplus} T_{\oplus} & \cos\chi\end{pmatrix},
\end{align}
where the angle $\chi$ is the colatitude of the laboratory,
which is $\chi \simeq 42.5^{\circ}$ for MEG
and $\chi\simeq 52.6^{\circ}$ for \babar.
The laboratory sidereal time $T_\oplus \equiv T - T_0$ 
is shifted relative to $T$
\cite{offset},
with $T_0 \simeq 3.9$~h for MEG
and $T_0 \simeq 12.5$~h for \babar. 
Neither the MEG nor the \babar detector frames 
coincide with the standard laboratory frame, 
so matching requires an extra rotation of $\psi$ about the $z$ axis
followed by an improper rotation $(x,y,z) \rightarrow (-x, z, +y)$,
\begin{align}
\label{netrot}
\mathcal{R_{\text{detector}}}  = 
\begin{pmatrix} 
-1 & 0 & 0 \\ 
0 & 0 & 1 \\ 
0 & +1 & 0 \end{pmatrix}
\begin{pmatrix} 
\cos\psi & \sin\psi & 0 \\ 
-\sin\psi& \cos\psi & 0 \\ 
0 & 0 & 1 \end{pmatrix} ,
\end{align}
where $\psi \simeq -30^{\circ}$ for MEG
and $\psi\simeq -50^{\circ}$ for \babar.

\renewcommand\arraystretch{0.8}
\begin{table}
\caption{Constraints deduced from Ref.~\cite{MEGlimit}.
\label{tablemeg}}
\centering
\setlength{\tabcolsep}{2pt}
\begin{tabular}{c c}
\hline\hline
Coefficients & Constraint (GeV$^{-1}$)\\ 
\hline
$	(m_{F}^{(5)})^{TJ}_{\mu e}, (m_{F}^{(5)})^{JZ}_{\mu e}, (m_{5F}^{(5)})^{TJ}_{\mu e}, (m_{5F}^{(5)})^{JZ}_{\mu e},	$&$	< 6.0\times 10^{-13}	$	\\
$	(a_{F}^{(5)})^{TTJ}_{\mu e}, (a_{F}^{(5)})^{TJZ}_{\mu e}, (a_{F}^{(5)})^{JTJ}_{\mu e}, (a_{F}^{(5)})^{JTK}_{\mu e},	$&$		$	\\
$	(a_{F}^{(5)})^{JJZ}_{\mu e}, (a_{F}^{(5)})^{JKZ}_{\mu e}, (a_{F}^{(5)})^{ZTJ}_{\mu e}, (a_{F}^{(5)})^{ZJZ}_{\mu e},	$&$		$	\\
$	(b_{F}^{(5)})^{TTJ}_{\mu e}$, $(b_{F}^{(5)})^{TJZ}_{\mu e},(b_{F}^{(5)})^{JTJ}_{\mu e}, (b_{F}^{(5)})^{JTK}_{\mu e},	$&$		$	\\
$	(b_{F}^{(5)})^{JJZ}_{\mu e}, (b_{F}^{(5)})^{JKZ}_{\mu e},(b_{F}^{(5)})^{ZTJ}_{\mu e}, (b_{F}^{(5)})^{ZJZ}_{\mu e},	$&$		$	\\
$	(H_{F}^{(5)})^{TJTJ}_{\mu e},(H_{F}^{(5)})^{TJTK}_{\mu e},(H_{F}^{(5)})^{TJJZ}_{\mu e},	$&$		$	\\
$	(H_{F}^{(5)})^{TJKZ}_{\mu e},(H_{F}^{(5)})^{TZTJ}_{\mu e},(H_{F}^{(5)})^{TZJZ}_{\mu e},	$&$		$	\\
$	(H_{F}^{(5)})^{JKTJ}_{\mu e},(H_{F}^{(5)})^{JKJZ}_{\mu e},(H_{F}^{(5)})^{JZTJ}_{\mu e},	$&$		$	\\
$	(H_{F}^{(5)})^{JZTK}_{\mu e},(H_{F}^{(5)})^{JZJZ}_{\mu e},(H_{F}^{(5)})^{JZKZ}_{\mu e}	$&$		$	\\[4pt]
$	(m_{F}^{(5)})^{TZ}_{\mu e},(m_{F}^{(5)})^{JK}_{\mu e},(m_{5F}^{(5)})^{TZ}_{\mu e},(m_{5F}^{(5)})^{JK}_{\mu e},	$&$	< 6.4\times 10^{-13}	$	\\
$	(a_{F}^{(5)})^{TTZ}_{\mu e},(a_{F}^{(5)})^{TJK}_{\mu e},(a_{F}^{(5)})^{JTZ}_{\mu e}, (a_{F}^{(5)})^{JJK}_{\mu e},	$&$		$	\\
$	(a_{F}^{(5)})^{ZTZ}_{\mu e}, (a_{F}^{(5)})^{ZJK}_{\mu e},(b_{F}^{(5)})^{TTZ}_{\mu e}, (b_{F}^{(5)})^{TJK}_{\mu e},	$&$		$	\\
$	(b_{F}^{(5)})^{JTZ}_{\mu e}, (b_{F}^{(5)})^{JJK}_{\mu e},(b_{F}^{(5)})^{ZTZ}_{\mu e}, (b_{F}^{(5)})^{ZJK}_{\mu e},	$&$		$	\\
$	(H_{F}^{(5)})^{TJTZ}_{\mu e},(H_{F}^{(5)})^{TZTZ}_{\mu e},(H_{F}^{(5)})^{JKTZ}_{\mu e},	$&$		$	\\
$	(H_{F}^{(5)})^{JZTZ}_{\mu e},(H_{F}^{(5)})^{TJJK}_{\mu e},(H_{F}^{(5)})^{TZJK}_{\mu e},	$&$		$	\\
$	(H_{F}^{(5)})^{JKJK}_{\mu e},(H_{F}^{(5)})^{JZJK}_{\mu e}	$&$		$	\\
\hline\hline 
\end{tabular}
\end{table}

\renewcommand\arraystretch{0.8}
\begin{table}
\caption{Constraints deduced from Ref.~\cite{BABARlimit}.
\label{tablebabar}}
\centering
\setlength{\tabcolsep}{2pt}
\begin{tabular}{c c}
\hline\hline
Coefficients & Constraint (GeV$^{-1}$)\\ 
\hline
$	(m_{F}^{(5)})^{TJ}_{\tau\mu},(m_{F}^{(5)})^{TZ}_{\tau\mu},(m_{F}^{(5)})^{JK}_{\tau\mu},(m_{F}^{(5)})^{JZ}_{\tau\mu}	$&$	 < 1.8\times 10^{-9}	$	\\[4pt]
$	(m_{5F}^{(5)})^{TJ}_{\tau\mu},(m_{5F}^{(5)})^{TZ}_{\tau\mu},(m_{5F}^{(5)})^{JK}_{\tau\mu},(m_{5F}^{(5)})^{JZ}_{\tau\mu}	$&$	< 2.0\times 10^{-9}	$	\\[4pt]
$	(a_{F}^{(5)})^{TTJ}_{\tau \mu},(a_{F}^{(5)})^{TTZ}_{\tau \mu},(a_{F}^{(5)})^{TJK}_{\tau \mu},(a_{F}^{(5)})^{TJZ}_{\tau \mu},	$&$	< 2.2\times 10^{-9}	$	\\
$	(b_{F}^{(5)})^{JTJ}_{\tau \mu},(b_{F}^{(5)})^{JTK}_{\tau \mu},(b_{F}^{(5)})^{JTZ}_{\tau \mu},(b_{F}^{(5)})^{JJK}_{\tau \mu},	$&$		$	\\
$	(b_{F}^{(5)})^{JJZ}_{\tau \mu},(b_{F}^{(5)})^{JKZ}_{\tau \mu},(b_{F}^{(5)})^{ZTJ}_{\tau \mu},(b_{F}^{(5)})^{ZTZ}_{\tau \mu}, 	$&$		$	\\
$	(b_{F}^{(5)})^{ZJK}_{\tau \mu},(b_{F}^{(5)})^{ZJZ}_{\tau \mu},(H_{F}^{(5)})^{JKTJ}_{\tau \mu},	$&$		$	\\
$	(H_{F}^{(5)})^{JKJZ}_{\tau \mu},(H_{F}^{(5)})^{JZTJ}_{\tau \mu},(H_{F}^{(5)})^{JZTK}_{\tau \mu},	$&$		$	\\
$	(H_{F}^{(5)})^{JZJZ}_{\tau \mu},(H_{F}^{(5)})^{JZKZ}_{\tau \mu},(H_{F}^{(5)})^{JZTZ}_{\tau \mu},	$&$		$	\\
$	(H_{F}^{(5)})^{JZJK}_{\tau \mu},(H_{F}^{(5)})^{JKTZ}_{\tau \mu},(H_{F}^{(5)})^{JKJK}_{\tau \mu}	$&$		$	\\[4pt]
$	(a_{F}^{(5)})^{JTJ}_{\tau \mu},(a_{F}^{(5)})^{JTK}_{\tau \mu},(a_{F}^{(5)})^{JTZ}_{\tau \mu},(a_{F}^{(5)})^{JJK}_{\tau \mu},	$&$	< 2.5\times 10^{-9}	$	\\
$	(a_{F}^{(5)})^{JJZ}_{\tau \mu},(a_{F}^{(5)})^{JKZ}_{\tau \mu},(a_{F}^{(5)})^{ZTJ}_{\tau \mu},(a_{F}^{(5)})^{ZTZ}_{\tau \mu},	$&$		$	\\
$	(a_{F}^{(5)})^{ZJK}_{\tau \mu},(a_{F}^{(5)})^{ZJZ}_{\tau \mu},(b_{F}^{(5)})^{TTJ}_{\tau \mu},(b_{F}^{(5)})^{TTZ}_{\tau \mu},	$&$		$	\\
$	(b_{F}^{(5)})^{TJK}_{\tau \mu},(b_{F}^{(5)})^{TJZ}_{\tau \mu},(H_{F}^{(5)})^{TJTJ}_{\tau \mu},	$&$		$	\\
$	(H_{F}^{(5)})^{TJTK}_{\tau \mu},(H_{F}^{(5)})^{TJJZ}_{\tau \mu},(H_{F}^{(5)})^{TJKZ}_{\tau \mu},	$&$		$	\\
$	(H_{F}^{(5)})^{TZTJ}_{\tau \mu},(H_{F}^{(5)})^{TZJZ}_{\tau \mu},(H_{F}^{(5)})^{TJTZ}_{\tau \mu},	$&$		$	\\
$	(H_{F}^{(5)})^{TJJK}_{\tau \mu},(H_{F}^{(5)})^{TZTZ}_{\tau \mu},(H_{F}^{(5)})^{TZJK}_{\tau \mu}	$&$		$	\\[4pt]
$	(m_{F}^{(5)})^{TJ}_{\tau e},(m_{F}^{(5)})^{TZ}_{\tau e},(m_{F}^{(5)})^{JK}_{\tau e},(m_{F}^{(5)})^{JZ}_{\tau e},	$&$	< 1.6\times 10^{-9}	$	\\
$	(m_{5F}^{(5)})^{TJ}_{\tau e},(m_{5F}^{(5)})^{TZ}_{\tau e},(m_{5F}^{(5)})^{JK}_{\tau e},(m_{5F}^{(5)})^{JZ}_{\tau e}	$&$		$	\\[4pt]
$	(a_{F}^{(5)})^{TTJ}_{\tau e},(a_{F}^{(5)})^{TTZ}_{\tau e},(a_{F}^{(5)})^{TJK}_{\tau e},(a_{F}^{(5)})^{TJZ}_{\tau e}	$&$	< 1.9\times 10^{-9}	$	\\[4pt]
$	(a_{F}^{(5)})^{JTJ}_{\tau e},(a_{F}^{(5)})^{JTK}_{\tau e},(a_{F}^{(5)})^{JTZ}_{\tau e},(a_{F}^{(5)})^{JJK}_{\tau e},	$&$	< 2.1\times 10^{-9}	$	\\
$	(a_{F}^{(5)})^{JJZ}_{\tau e},(a_{F}^{(5)})^{JKZ}_{\tau e},(a_{F}^{(5)})^{ZTJ}_{\tau e},(a_{F}^{(5)})^{ZTZ}_{\tau e},	$&$		$	\\
$	(a_{F}^{(5)})^{ZJK}_{\tau e},(a_{F}^{(5)})^{ZJZ}_{\tau e}	$&$		$	\\[4pt]
$	(b_{F}^{(5)})^{TTJ}_{\tau e},(b_{F}^{(5)})^{TTZ}_{\tau e},(b_{F}^{(5)})^{TJK}_{\tau e},(b_{F}^{(5)})^{TJZ}_{\tau e},	$&$	< 2.0\times 10^{-9}	$	\\
$	(b_{F}^{(5)})^{JTJ}_{\tau e},(b_{F}^{(5)})^{JTK}_{\tau e},(b_{F}^{(5)})^{JTZ}_{\tau e},(b_{F}^{(5)})^{JJK}_{\tau e},	$&$		$	\\
$	(b_{F}^{(5)})^{JJZ}_{\tau e},(b_{F}^{(5)})^{JKZ}_{\tau e},(b_{F}^{(5)})^{ZTJ}_{\tau e},(b_{F}^{(5)})^{ZTZ}_{\tau e},	$&$		$	\\
$	(b_{F}^{(5)})^{ZJK}_{\tau e},(b_{F}^{(5)})^{ZJZ}_{\tau e},(H_{F}^{(5)})^{TJTJ}_{\tau e},	$&$		$	\\
$	(H_{F}^{(5)})^{TJTK}_{\tau e},(H_{F}^{(5)})^{TJJZ}_{\tau e},(H_{F}^{(5)})^{TJKZ}_{\tau e},	$&$		$	\\
$	(H_{F}^{(5)})^{TZTJ}_{\tau e},(H_{F}^{(5)})^{TZJZ}_{\tau e},(H_{F}^{(5)})^{TJTZ}_{\tau e},	$&$		$	\\
$	(H_{F}^{(5)})^{TJJK}_{\tau e},(H_{F}^{(5)})^{TZTZ}_{\tau e},(H_{F}^{(5)})^{TZJK}_{\tau e},	$&$		$	\\
$	(H_{F}^{(5)})^{JKTJ}_{\tau e},(H_{F}^{(5)})^{JKJZ}_{\tau e},(H_{F}^{(5)})^{JZTJ}_{\tau e},	$&$		$	\\
$	(H_{F}^{(5)})^{JZTK}_{\tau e},(H_{F}^{(5)})^{JZJZ}_{\tau e},(H_{F}^{(5)})^{JZKZ}_{\tau e},	$&$		$	\\
$	(H_{F}^{(5)})^{JZTZ}_{\tau e},(H_{F}^{(5)})^{JZJK}_{\tau e},(H_{F}^{(5)})^{JKTZ}_{\tau e},	$&$		$	\\
$	(H_{F}^{(5)})^{JKJK}_{\tau e}	$&$		$	\\
\hline\hline 
\end{tabular}
\end{table}

Experiments searching for LV 
aim to measure the coefficients for LV in the SCF.
The above transformations show that the experimental observables
in the detector frame are functions of $\chi, \psi$, $T_\oplus$,
and the SCF coefficients
and that a given coefficient with $n$ Lorentz indices
is generically accompanied by oscillations in $T_{\oplus}$
involving from zero to $n$ harmonics.
As a result,
data taken with time stamps can be binned in sidereal time
and used to extract the amplitudes and phases of the various harmonics,
yielding a series of independent constraints on the SCF coefficients.
The dependence on colatitude and longitude implies 
that different experiments are sensitive 
to distinct coefficient combinations.

For the MEG and \babar experiments,
the published limits on the branching ratios
can be viewed as time-averaged measurements.
The time averages of the results \rf{decayrate} and \rf{decayratetau}
involve only rotation-invariant combinations
of the SCF coefficients appearing in Table \ref{operators},
although they can still depend on the experiment colatitude $\chi$
and detector orientation $\psi$.
Note also that the integration \rf{decayratetau}
over the full final-state phase space 
means that the decay rate for \babar
is unaffected by the rotation to the SCF. 
The three published experimental limits 
\cite{MEGlimit,BABARlimit}
on the decays yield three constraints 
on the combinations of SME coefficients in the SCF
given in Eqs.\ \rf{decayrate} and \rf{decayratetau}.
Calculation reveals that all types of independent coefficients
in Table \ref{operators}
contribute to the time-averaged signals.
Following standard procedure in the field
\cite{tables},
we can transform the three experimental constraints
into limits on independent coefficient components taken one at a time.
This procedure yields the 576 constraints on LV 
presented in Tables \ref{tablemeg} and \ref{tablebabar}.
Of these, 
the entries involving the coefficients 
$(a_{F}^{(5)})^{\mu\alpha\beta}_{AB}$ and
$(b_{F}^{(5)})^{\mu\alpha\beta}_{AB}$
also are bounds on CPT violation. 
Each entry is a constraint at the 90\% confidence level 
on the modulus of the real and imaginary parts
of a coefficient component in the SCF,
with indices $J$ and $K\neq J$ taking the values $X$ or $Y$.

To summarize,
an analysis of published data from the MEG and \babar experiments
places constraints on 576 independent coefficients 
for electromagnetic flavor-changing Lorentz and CPT violation 
in the charged-lepton sector.
The results are consistent with no flavor-changing Lorentz violation
in the range of parts in $10^{-13}$ GeV$^{-1}$ to $10^{-9}$ GeV$^{-1}$,
and they establish a bar excluding flavor-changing LV effects on these scales.
Excellent prospects exist for future improvements on these results,
both via analyses incorporating sidereal and annual time variations 
and via increased sensitivities 
to $\ell_A \rightarrow \ell_B + \gamma$ and related decays 
in upcoming experiments
\cite{MEGII,BELLEII,mu2e,comet,mu3e}.

This work is supported in part by 
the U.S.\ Department of Energy 
under grants {DE}-SC0010120 and {DE}-AC05-06OR23177,
by the U.S.\ National Science Foundation 
under grants PHY-1748958 and PHY-2013184,
by the U.K.\ Science and Technology Facilities Council
under grant ST/T006048/1,
and by the Indiana University Center for Spacetime Symmetries.


\begin{thebibliography}{}

\bibitem{wu57}
C.S.\ Wu, E.\ Ambler, R.W.\ Hayward, D.D.\ Hoppes, and R.P.\ Hudson,
Phys.\ Rev.\ {\bf 105}, 1413 (1957).

\bibitem{jc64}
J.H.\ Christenson, J.W.\ Cronin, V.L.\ Fitch, and R.\ Turlay,
Phys.\ Rev.\ Lett.\ {\bf 13}, 138 (1964).

\bibitem{rd68}
R.\ Davis, Jr., D.S.\ Harmer, and K.C.\ Hoffman,
Phys.\ Rev.\ Lett.\ {\bf 20}, 1205 (1968);
Q.R.\ Ahmad \etal,
Phys.\ Rev.\ Lett.\ {\bf 87}, 071301 (2001).

\bibitem{sk98}
Y.\ Fukuda \etal,
Phys.\ Rev.\ Lett.\ {\bf 81}, 1562 (1998).

\bibitem{ksp}
V.A.\ Kosteleck\'y and S.\ Samuel,
Phys.\ Rev.\ D {\bf 39}, 683 (1989);
V.A.\ Kosteleck\'y and R.\ Potting,
Nucl.\ Phys.\ B {\bf 359}, 545 (1991);
Phys.\ Rev.\ D {\bf 51}, 3923 (1995).

\bibitem{tables}
V.A.\ Kosteleck\'y and N.\ Russell,
{\it Data Tables for Lorentz and CPT Violation},
Rev. Mod. Phys. {\bf 83}, 11 (2011), 
arXiv:0801.0287v15 (2022).

\bibitem{ak98}
V.A.\ Kosteleck\'y,
Phys.\ Rev.\ Lett.\ {\bf 80}, 1818 (1998).

\bibitem{smdecays}
W.J.\ Marciano and A.I.\ Sanda,
Phys.\ Lett.\ B {\bf 67}, 303 (1977);
S.M.\ Bilenky, S.T.\ Petcov, and B.\ Pontecorvo,
Phys.\ Lett.\ B {\bf 67}, 309 (1977);
B.W.\ Lee, S.\ Pakvasa, R.E.\ Schrock, and H.\ Sugawara,
Phys.\ Rev.\ Lett.\ {\bf 38}, 937 (1977).

\bibitem{ck97}
D.\ Colladay and V.A.\ Kosteleck\'y,
Phys.\ Rev.\ D {\bf 55}, 6760 (1997);
Phys.\ Rev.\ D {\bf 58}, 116002 (1998).

\bibitem{ak04}
V.A.\ Kosteleck\'y,
Phys.\ Rev.\ D {\bf 69}, 105009 (2004).

\bibitem{review}
For reviews see,
for example,
A.\ Hees, Q.G.\ Bailey, A.\ Bourgoin, H.\ Pihan-Le Bars,
C.\ Guerlin, and C.\ Le Poncin-Lafitte, 
Universe {\bf 2}, 30 (2016);
J.D.\ Tasson,
Rep.\ Prog.\ Phys.\ {\bf 77}, 062901 (2014);
C.M.\ Will, 
Liv.\ Rev.\ Rel.\ {\bf 17}, 4 (2014);
R.\ Bluhm,
Lect.\ Notes Phys.\ {\bf 702}, 191 (2006).

\bibitem{owg}
O.W.\ Greenberg,
Phys.\ Rev.\ Lett.\ {\bf 89}, 231602 (2002). 

\bibitem{kt09}
V.A.\ Kosteleck\'y and J.D.\ Tasson,
Phys.\ Rev.\ Lett.\ {\bf 102}, 010402 (2009).

\bibitem{cg99}
S.R.\ Coleman and S.L.\ Glashow, 
Phys.\ Rev.\ D {\bf 59}, 116008 (1998).

\bibitem{yi03}
E.O.\ Iltan,
JHEP {\bf 06}, 016 (2003).

\bibitem{gh20}
Y.M.P.\ Gomes and J.A.\ Helayel-Neto,
Eur.\ Phys.\ J.\ C {\bf 80}, 287 (2020).

\bibitem{cks20}
A.\ Crivellin, F.\ Kirk and M.\ Schreck,
JHEP {\bf 04}, 082 (2021).

\bibitem{hp11}
S.\ Hollenberg and P.B.\ Pal,
Phys.\ Lett.\ B {\bf 701}, 89 (2011).

\bibitem{nowt16}
J.P.\ Noordmans, C.J.G.\ Onderwater, H.W.\ Wilschut, and R.G.E.\ Timmermans,
Phys.\ Rev.\ D {\bf 93}, 116001 (2016).

\bibitem{lmt16}
M.A.\ L\'opez-Osorio, E.\ Mart\'\i{}nez-Pascual, and J.J.\ Toscano,
J.\ Phys.\ G {\bf 43}, 025003 (2016).

\bibitem{msf17}
V.E.\ Mouchrek-Santos and M.M.\ Ferreira, Jr.,
Phys.\ Rev.\ D {\bf 95}, 071701 (2017).

\bibitem{gmn20}
Y.M.P.\ Gomes, P.C.\ Malta, and M.J.\ Neves,
Phys.\ Rev.\ D {\bf 101}, 075001 (2020).

\bibitem{kl19}
V.A.\ Kosteleck\'y and Z.\ Li,
Phys.\ Rev.\ D {\bf 99}, 056016 (2019);
Phys.\ Rev.\ D {\bf 103}, 024059 (2021).

\bibitem{cpt}
J.S.\ Bell, 
Proc.\ Roy.\ Soc.\ (London) {\bf A231}, 479 (1955);
W.\ Pauli, 
in W.\ Pauli, ed.,
{\it Niels Bohr and the Development of Physics},
McGraw-Hill, New York, 1955;
G.\ L\"uders, 
Ann.\ Phys.\ (N.Y.) {\bf 2}, 1 (1957).

\bibitem{nonpertmue}
W.~Dekens, E.E.~Jenkins, A.V.~Manohar and P.~Stoffer,
JHEP {\bf 01}, 088 (2019).

\bibitem{taudecays}
A.~Celis, V.~Cirigliano, and E.~Passemar,
Phys.\ Rev.\ D {\bf 89}, 095014 (2014).

\bibitem{EFTmatching}
E.E.~Jenkins, A.V.~Manohar and P.~Stoffer,
JHEP {\bf 3}, 016 (2018).

\bibitem{kl01}
V.A.\ Kosteleck\'y and R.\ Lehnert,
Phys.\ Rev.\ D {\bf 63}, 065008 (2001).  

\bibitem{MEGlimit}
A.M.~Baldini \etal,
Eur.\ Phys.\ J.\ C {\bf 76}, 434 (2016).

\bibitem{BABARlimit}
B.~Aubert \etal,
Phys.\ Rev.\ Lett.\ {\bf 104}, 021802 (2010).

\bibitem{MEGdetector}
J.~Adam \etal,
Eur.\ Phys.\ J.\ C {\bf 73}, 2365 (2013).

\bibitem{BABARreport}
D.~Boutigny \etal,
SLAC-R-0457 (1995).

\bibitem{babar17}
G.~Ciezarek, M.~Franco Sevilla, B.~Hamilton, R.~Kowalewski, 
T.~Kuhr, V.~L\"uth, and Y.~Sato,
Nature {\bf 546}, 227 (2017).

\bibitem{sunframe}
R.\ Bluhm, V.A.\ Kosteleck\'y, C.D.\ Lane, and N.\ Russell,
Phys.\ Rev.\ D {\bf 68}, 125008 (2003);
Phys.\ Rev.\ Lett.\ {\bf 88}, 090801 (2002);
V.A.\ Kosteleck\'y and M.\ Mewes,
Phys.\ Rev.\ D {\bf 66}, 056005 (2002).  

\bibitem{offset}
V.A.\ Kosteleck\'y, A.C.\ Melissinos, and M.\ Mewes,
Phys.\ Lett.\ B {\bf 761}, 1 (2016);
Y.\ Ding and V.A.\ Kosteleck\'y,
Phys.\ Rev.\ D {\bf 94}, 056008 (2016).

\bibitem{MEGII}
A.M.~Baldini \etal,
Eur.\ Phys.\ J.\ C {\bf 78}, 380 (2018).

\bibitem{BELLEII}
E.~Kou \etal,
PTEP {\bf 2019}, 123C01 (2019).

\bibitem{mu2e}
L.~Bartoszek \etal,
arXiv:1501.05241.

\bibitem{comet}
R.~Abramishvili \etal,
PTEP {\bf 2020}, 033C01 (2020).

\bibitem{mu3e}
A.~Blondel \etal,
arXiv:1301.6113.

\end{thebibliography}
\end{document}